\documentclass[fleqn,twoside]{article}
\usepackage[headings]{espcrc2}
\usepackage{amsmath}
\usepackage{amssymb}
\usepackage{epsfig}
\usepackage{euscript}
\usepackage{fancybox}
\usepackage{color}
\def\mathswitchr#1{\relax\ifmmode{\mathrm{#1}}\else$\mathrm{#1}$\fi}

\newcommand {\pslash}{\hbox{$\not\hbox{\kern-2.3pt $p$}$}}

\usepackage{cite}

\usepackage{epic}
\def\alf1{ {\alpha\over\pi} }
\def\rQCD{{\rm QCD}}
\def\rQCED{{\rm QCED}}
\readRCS
$Id: espcrc2.tex,v 1.2 2004/02/24 11:22:11 spepping Exp $
\usepackage{graphicx}
\usepackage[figuresright]{rotating}

{\small
\title{
BU-HEPP-06-04\\
Jan., 2006\\
~~~~\\
~~~~\\
New Applications of Resummation in Non-Abelian Gauge Theories:
 QED$\otimes$QCD Exponentiation for LHC Physics, IR-Improved DGLAP Theory 
 and Resummed Quantum Gravity}

\author{
        B.F.L. Ward\address{Physics Department, Baylor University, \\
        Waco, Texas, USA}%
        \thanks{Work supported in part by US DOE grant DE-FG02-05ER42399 and by           NATO grant PST.CLG.980342}}
\runauthor{B.F.L. Ward}
}
\begin{document}
{\small
\begin{abstract}
We present the elements of three applications of resummation methods
in non-Abelian gauge theories: (1), $QED\otimes QCD$ exponentiation
and shower/ME matching for LHC physics; 
(2), IR improvement of DGLAP theory; (3), resummed
quantum gravity and the final state of Hawking radiation.
In all cases, the extension of the YFS approach, originally introduced
for Abelian gauge theory, to non-Abelian gauge theories, 
QCD and quantum general relativity, leads to new results and solutions which we briefly summarize.
%\vspace{1cm}
%\centerline{BU-HEPP-06-04, Jan., 2006}
\vspace{1pc}
\end{abstract}
%\centerline{BU-HEPP-06-04, Jan., 2006}
}
\maketitle
%\centerline{BU-HEPP-06-04, Jan., 2006}
%\section{FORMAT}
{\small
\section{Introduction}
\label{intro}
Multiple gluon (n(G)) effects are already needed in many high energy
collider physics scenarios, such as t\=t production
at FNAL, polarized pp processes at RHIC, b\=b production at FNAL,$\cdots$,
and contribute a large part of the current uncertainty on $m_t$
,$\sim$ 2-3 GeV~\cite{grannis}. For the LHC, and any
TeV scale linear collider, the rather more demanding
requirements make exponentiated soft n(G) results, {\em exact} 
through ${\cal O}(\alpha_s^2)$, realized by MC methods in the 
presence of parton showers without double counting on an event-by-event basis
with {\em exact} phase space, an essential part of the necessary theory.\par

For such precision QCD results,$\sim 1\%$, may
QED higher order corrections also be relevant?
The standard treatments of resummation
in the two theories treat them separately, where we have considerable
literature on resummation in QCD~\cite{sterm,cattrent,qcdref,delaney} and on exponentiation in QED~\cite{jack-schar,yfs,jadward,yellowbook}. We use the words resummation and exponentiation interchangeably here because in all 
cases a leading exponential factor is
identified in the resummations which we discuss in this paper. 
Results from Refs.~\cite{cern2000,spies,james1,roth,james2} 
show a few per mille level QED effect from structure function evolution 
when QED and QCD DGLAP~\cite{dglap} kernels are treated simultaneously.
After reviewing the YFS theory and our extension of it to QCD in
the next Section, in Section~\ref{qcdqedexp} we
combine~\cite{qcdqed} the respective two exponential algebras at the level of the amplitudes to treat QCD and QED 
exponentiation simultaneously, QED$\otimes$QCD exponentiation, to be able to 
isolate, on an event-by-event basis,
the possible interplay between the large effects from soft gluons and soft
photons.
% in the threshold region of heavy particle production at the LHC
%where soft QED effects, as it is well-known, may be large.
We also show that the QCD exponentiation extends naturally to quantum
gravity and allows us to develop UV finite 
resummed quantum gravity~\cite{bw1}.
%where it can be realized
%that ours is a new version
%of the resummation approach among the four general
%approaches to quantum general relativity discussed in Ref.~\cite{wein1}.
We find surprisingly
that 
%without modifying Einstein's theory, 
%when we apply our non-Abelian
our extension of the ideas of Ref.~\cite{yfs}
%Yennie, Frautschi and Suura 
to Feynman's
formulation~\cite{ff1,ff2} of Einstein's theory, the UV behavior of the respective theory becomes convergent.
\par
 
The size of the QED$\otimes$QCD threshold effects is illustrated
in Sect.~\ref{smematch}, wherein
%where
%As a natural application of the theory, 
we present 
%in Sect.~\ref{smematch} 
as well a new approach to shower/ME
matching~\cite{frix-web} for exact fixed-higher-order results.
We also present IR-improved DGLAP theory~\cite{irdglap}
in this Section.
%that 
%we
%thereby make contact with the results on 
%Refs.~\cite{cern2000,spies,james1,roth,james2}
%both from the standpoint of the type of QED effect we find and from the
%standpoint that 
%our resummation algebra allows us to improve
%the DGLAP QCD kernels as well and open up the road to
%an improved QCD evolution for any given level of fixed order 
%exactness for these kernels. We refer to this as IR-improved 
%DGLAP theory~\cite{irdglap}. 
In Sect.~\ref{hawkrad} we discuss 
%illustrate
%the application of resummed quantum gravity to 
the final state
of Hawking~\cite{hawk1} radiation for an originally 
very massive black hole in resummed quantum gravity. 
%Sect.~\ref{conclsn} contains summary remarks. 
\par

%Success with the extension of the YFS-style of resummation, in which
%the Feynman series is re-arranged, nothing is dropped and  exactness is
%maintained throughout, leads us naturally to apply it to the other 
%outstanding problem in the application of quantum field theory to 
%phenomenology, that of the problem of quantum general relativity, where
%the four general types of approaches to the problem can be found summarized
%in Ref.~\cite{wein1}. It can be realized that ours~\cite{bw1,bw2,bw3,bw4} is a new version
%of the resummation approach noted in the latter reference. We find surprisingly
%that, without modifying Einstein's theory, when we apply our non-Abelian
%extension of the ideas of Yennie, Frautschi and Suura to Feynman's
%formulation~\cite{ff1,ff2} of Einstein's theory, the UV behavior of the respective theory becomes convergent. We illustrate an application of this result 
%to the final
%state of Hawking~\cite{hawk1} radiation for an originally 
%very massive black hole.\par

%Our paper proceeds as follows. In the next Section, we review
%the YFS theory and its extension to QCD. In Sect.~\ref{qcdqedexp} 
%we present the further extension to QED$\otimes$QCD exponentiation
%and to quantum general relativity. In Sect.~\ref{smematch} we 
%discuss QED$\otimes$QCD exponentiation of threshold effects, shower/ME matching
%and IR-improved DGLAP theory at the LHC. In Sect.~\ref{hawkrad} we
%present the final state of Hawking radiation in resummed quantum gravity.
%Sect.~\ref{conclsn} contains summary remarks.\par

\section{Review of YFS Theory and Its Extension to QCD}
\label{rev-qed-qcd}
We consider first the QED case presented in Refs.~\cite{sjward} and realized
by MC methods, for which, as we illustrate here for $e^+(p_1)e^-(q_1)\rightarrow \bar{f}(p_2) f(q_2) +n(\gamma)(k_1,\cdot,k_n)$,
renormalization group improved YFS theory~\cite{bflwyfs}
% (PRD{\bf 36}(1987)939) 
gives{\small
\begin{equation}
\begin{split}
%\begin{eqnarray}
d\sigma_{exp}=e^{2\alpha\,Re\,B+2\alpha\,
\tilde B}\sum_{n=0}^\infty\frac{1}{n!}\int\prod_{j=1}^n\frac{d^3k_j}{k_j^0
}\cr \qquad\int \frac{d^4y}{(2\pi)^4}e^{iy(p_1+q_1-p_2-q_2-\sum_jk_j)+D}
\cr \qquad\bar\beta_n(k_1,\dots,k_n)\frac{d^3p_2d^3q_2}{p_2^0q_2^0}
\label{eqone}\end{split}\end{equation}}
%\end{eqnarray}}
%\end{equation}}
where the YFS real and virtual infrared functions $\tilde B,~D,~B$
%real infrared function $\tilde B$ and the virtual infrared function $B$ 
are known~\cite{yfs,sjward}. 
%%and where we note the usual connections{\small
%%\[2\alpha\,\tilde B = \int^{k\le K_{max}}{d^3k\over k_0}\tilde S(k)\nonumber\]
%%\begin{equation}D=\int d^3k{\tilde S(k)\over k^0}\left(e^{-iy\cdot k}-\theta(K_{
%%max}-k)\right)\label{eqtwo}\end{equation}} for the standard YFS 
%%infrared emission factor{\small
%%\begin{equation}\tilde S(k)= {\alpha\over4\pi^2}\left[Q_fQ_{
%%{\llap{\phantom f}^{\sstl(}\bar f^{\sstl^)}{}}'}\left({p_1\over p_1\cdot k}-{q_1
%%\over q_1\cdot k}\right)^2+(\dots)\right]\label{eqthree}\end{equation}} 
%%if $Q_f$
%%is the electric charge of $f$ in units of the positron charge. 
Examples of the YFS hard photon residuals $\bar\beta_i$ in (\ref{eqone}), 
$i=0,1,2$,
are given in the fourth paper in 
Refs.~\cite{sjward} for the MC  BHLUMI 4.04. 
%and yield the
%{\Color{Red}S. Jadach {\it et al.},CPC{\bf 102}(1997)229} for BHLUMI 4.04 {\Color{Black}$\Rightarrow$}
%YFS exponentiated exact ${\cal O}(\alpha)$
%and LL ${\cal O}(\alpha^2)$
%cross section for Bhabha scattering via a
%corresponding Monte Carlo realization
%of (\ref{eqone}).
\par

In Ref.~\cite{qcdref}
%hep-ph/0210357(ICHEP02), Acta Phys.Polon.B33,1543-1558,2002,
we have extended this YFS theory to QCD:{\small
\begin{equation}
\begin{split}
d\hat\sigma_{\rm exp}&= \sum_n d\hat\sigma^n \\
         &=e^{\rm SUM_{IR}(QCD)}\sum_{n=0}^\infty\int\prod_{j=1}^n\frac{d^3
k_j}{k_j}\int\frac{d^4y}{(2\pi)^4}\\
&e^{iy\cdot(P_1+P_2-Q_1-Q_2-\sum k_j)+
D_\rQCD}\\
&*\tilde{\bar\beta}_n(k_1,\ldots,k_n)\frac{d^3P_2}{P_2^{\,0}}\frac{d^3Q_2}{Q_2^{\,0}}
\end{split}
\label{subp15a}
\end{equation}}
where now the hard gluon residuals 
$\tilde{\bar\beta}_n(k_1,\ldots,k_n)$
represented by{\small 
\[\tilde{\bar\beta}_n(k_1,\ldots,k_n)= \sum_{\ell=0}^\infty 
\tilde{\bar\beta}^{(\ell)}_n(k_1,\ldots,k_n)\]}
are free of all infrared divergences to all 
orders in $\alpha_s(Q)$. The functions 
${\rm SUM_{IR}(QCD)},~D_\rQCD,~\tilde{\bar\beta}_n(k_1,\ldots,k_n)$
are the QCD analoga of the QED functions $2\alpha\,Re\,B+2\alpha\,
\tilde B,~D,~\bar\beta_n(k_1,\ldots,k_n)$ and are defined in Ref.~\cite{qcdref,delaney}. 
%Note that in defining the QCD IR and hard residuals, one can allow
%the running of $\alpha_s$ in the respective integrations to resum 
%a larger class of higher order effects as usual.  
For reference, the respective
process exponentiated in (\ref{subp15a}) can be taken to be
$q+\bar{q}'\rightarrow q''+\bar{q}'''+n(G)$ for quarks $q,q''$ and 
anti-quarks $\bar{q}',\bar{q}'''$.
\par
%We stress that the arguments
%in the earlier papers in Ref.~\cite{delaney} 
%({\Color{Red}DeLaney {\it et al.} PRD{\bf 52}(1995)108, 
%PLB{\bf 342}(1995)239}) 
%are not really sufficient to derive the respective 
%analog of eq.(\ref{subp15a});
%for, they did not really expose the compensation between
%the left over genuine non-Abelian IR virtual and real singularities
%between $\int dPh\bar\beta_n$ and $\int dPh\bar\beta_{n+1}$ respectively
%that really allows us to isolate $\tilde{\bar\beta}_j$ and distinguishes
%QCD from QED, where no such compensation occurs.
Our exponential factor corresponds to the $N=1$ 
term in the exponent
in Gatherall's formula~\cite{gatherall} 
%{\Color{PineGreen}(Phys. Lett.{\bf B133}(1983)90)} 
for the general exponentiation of the eikonal cross sections for non-Abelian gauge theory; his result
is an approximate one in which everything that does 
not eikonalize and exponentiate is dropped whereas 
our result (\ref{subp15a}) is exact.\par

\section{Extension to QED$\otimes$QCD and Quantum Gravity} 
\label{qcdqedexp}
Simultaneous exponentiation of
QED and QCD higher order effects~\cite{qcdqed}
%%hep-ph/0404087}
gives{\small
\begin{eqnarray}
B^{nls}_{QCD} \rightarrow B^{nls}_{QCD}+B^{nls}_{QED}\equiv B^{nls}_{QCED},\cr
{\tilde B}^{nls}_{QCD}\rightarrow {\tilde B}^{nls}_{QCD}+{\tilde B}^{nls}_{QED}\equiv {\tilde B}^{nls}_{QCED}, \cr
%%{\tilde S}^{nls}_{QCD}\rightarrow {\tilde S}^{nls}_{QCD}+{\tilde S}^{nls}_{QED}\equiv {\tilde S}^{nls}_{QCED}
\end{eqnarray}} 
which leads to{\small
\begin{eqnarray}
%\begin{split}
d\hat\sigma_{\rm exp} = e^{\rm SUM_{IR}(QCED)}
   \sum_{{n,m}=0}^\infty\int\prod_{j_1=1}^n\frac{d^3k_{j_1}}{k_{j_1}} 
\prod_{j_2=1}^m\frac{d^3{k'}_{j_2}}{{k'}_{j_2}}\cr
\int\frac{d^4y}{(2\pi)^4}e^{iy\cdot(p_1+q_1-p_2-q_2-\sum k_{j_1}-\sum {k'}_{j_2})+
D_\rQCED} \cr
\tilde{\bar\beta}_{n,m}(k_1,\ldots,k_n;k'_1,\ldots,k'_m)\frac{d^3p_2}{p_2^{\,0}}\frac{d^3q_2}{q_2^{\,0}},\cr
%\end{split}
\label{subp15b}
\end{eqnarray}}
where  the new YFS residuals 
$\tilde{\bar\beta}_{n,m}(k_1,\ldots,k_n;k'_1,\ldots,k'_m)$, with $n$ hard gluons and $m$ hard photons,
represent the successive application
of the YFS expansion first for QCD and subsequently for QED. 

The infrared functions are now{\small 
\begin{eqnarray}
{\rm SUM_{IR}(QCED)}&=&2\alpha_s\Re B^{nls}_{QCED}+2\alpha_s{\tilde B}^{nls}_{QCED}\cr
%%D_\rQCED=\int \frac{dk}{k^0}\left(e^{-iky}-\theta(K_{max}-k^0)\right){\tilde S}^{nls}_{QCED}
D_\rQCED &=& D_\rQCD+D.
\label{irfns}
\end{eqnarray}}
%where the implicit $K_{max}$ is the same dummy parameter for QCD and QED.

We have a new IR Algebra(QCED) which is characterized by the following
estimates of the respective average values of Bjorken's variable:~{\small
$x_{avg}(QED)\cong \gamma(QED)/(1+\gamma(QED))$,
$x_{avg}(QCD)\cong \gamma(QCD)/(1+\gamma(QCD))$, with
$\gamma(A)=\frac{2\alpha_{A}{\cal C}_A}{\pi}(L_s
-1)$, $A=QED,QCD$,
${\cal C}_A=Q_f^2, C_F$, respectively, for 
$A=QED,QCD$, where ~$L_s$ is the big log,}
so that QCD dominant corrections happen an
order of magnitude earlier than those for QED; 
%so that
the leading $\tilde{\bar\beta}_{0,0}^{(0,0)}$-level
gives a good estimate of the size of the effects we study. 
\par

We can apply 
(\ref{subp15b}) to quantum general relativity: For the scalar 2-point function
we get~\cite{bw1}{\small
\begin{equation}
i\Delta'_F(k)|_{\text{resummed}} =  \frac{ie^{B''_g(k)}}{(k^2-m^2-\Sigma'_s+i\epsilon)}
\label{resum}
\end{equation}}
for{\small
\begin{equation}
\begin{split} 
B''_g(k)&= -2i\kappa^2k^4\frac{\int d^4\ell}{16\pi^4}\frac{1}{\ell^2-\lambda^2+i\epsilon}\\
&\qquad\frac{1}{(\ell^2+2\ell k+\Delta +i\epsilon)^2}
\end{split}
\label{yfs1} 
\end{equation}}
This is the basic result.
\par
Note the following:
$\Sigma'_s$ starts in ${\cal O}(\kappa^2)$, 
so we may drop it in calculating one-loop effects;
explicit evaluation gives, for the deep UV regime,{\small
$B''_g(k) = \frac{\kappa^2|k^2|}{8\pi^2}\ln\left(\frac{m^2}{m^2+|k^2|}\right)$,
} so that (\ref{resum}) 
falls faster than any power of $|k^2|$; and,
if $m$ vanishes, using the usual $-\mu^2$ normalization point we get{\small\;
$B''_g(k)=\frac{\kappa^2|k^2|}{8\pi^2}
\ln\left(\frac{\mu^2}{|k^2|}\right)$}
so that (\ref{resum}) again vanishes faster than any power of $|k^2|$! 
These observations mean~\cite{bw1} that, 
when the respective analoga of (\ref{resum}) are used, one-loop corrections are finite.
Indeed, all quantum gravity loops are UV finite~\cite{bw1}.
%(MPL{\bf A17}(2002)2371)
We refer to this approach to quantum general relativity as resummed quantum
gravity (RQG).\par

\section{QED$\otimes$QCD Threshold Corrections, Shower/ME Matching and IR-Improved DGLAP Theory at the LHC}
\label{smematch}
We shall apply the new simultaneous QED$\otimes$QCD exponentiation
calculus to the single Z production with leptonic decay 
at the LHC ( and at FNAL)
to focus on the ISR alone, for definiteness.
See also the work of Refs.~\cite{baur,ditt,zyk}
%Baur {\it et al.}, Dittmaier and Kramer,
%Zykunov  
for exact ${\cal O}(\alpha)$ results and 
Refs.~\cite{van1,van2,anas}
%Hamberg {\it et al.}, van Neerven and Matsuura and
%Anastasiou {\it et al.} 
for exact ${\cal O}(\alpha_s^2)$ results.

For the basic formula{\small
\begin{equation}\begin{split}
d\sigma_{exp}(pp\rightarrow V+X\rightarrow \bar\ell \ell'+X') =\cr
\sum_{i,j}\int dx_idx_j F_i(x_i)F_j(x_j)d\hat\sigma_{exp}(x_ix_js),\cr
\label{sigtot}\end{split} 
\end{equation}}
we use the result in (\ref{subp15b}) here with semi-analytical
methods and structure functions from Ref.~\cite{mrst1}.
%Martin {\it et al.}
A MC realization will appear elsewhere~\cite{elsewh}.

Here we make the following observations.
\begin{itemize}
\item
In (\ref{sigtot})
we do not attempt at this time to replace 
Herwig~\cite{herwig} and/or Pythia~\cite{pythia} --
we intend to combine our exact YFS calculus, $d\hat\sigma_{exp}(x_ix_js)$, with Herwig and/or Pythia
by using them/it ``in lieu'' of $\{F_i\}$ as follows:
A. Use a Herwig/Pythia shower for $p_T\le \mu$, and YFS $nG$ radiation for $p_T> \mu$.\\
B. Or, expand the Herwig/Pythia shower formula $\otimes d\sigma_{exp}$ 
and adjust
$\tilde{\bar\beta}_{n,m}$ to exactness for the desired order with new
$\tilde{\bar\beta}_{n,m}'$.\\
In either A or B, we first use $\{F_i\}$ to pick $(x_1,x_2)$; make an event with $d\sigma_{exp}$; 
then shower event using Herwig/Pythia via Les Houches recipe~\cite{leshouches}.
\item This combination of theoretical constructs can be 
systematically improved with
exact results order-by-order in $\alpha_s,\alpha$, with exact phase space.
\item Possible new parton showers such as one based on the  recent alternative parton evolution algorithm in Refs.`\cite{jadskrz}
%Skrzypek,{\color{magenta} acta. phys. pol.{\bf b35}, 745 (2004)}}, 
can also be used.
\item Due to its lack of color coherence~\cite{mmm} Isajet~\cite{isajet} is not considered here.
\end{itemize}

With this said, we compute , with and without QED, the ratio
$r_{exp}=\sigma_{exp}/\sigma_{Born}$
to get the results
(We stress that we {\em do not} use the narrow resonance approximation here.)
{\small
\begin{equation}
r_{exp}=
\begin{cases}
1.1901&, \text{\small QCED}\equiv \text{\small QCD+QED,~~LHC}\\
1.1872&, \text{\small QCD,~~LHC}\\
1.1911&, \text{\small QCED}\equiv \text{\small QCD+QED,~~Tevatron}\\
1.1879&, \text{\small QCD,~~Tevatron.}\\
\end{cases}
\label{res1}
\end{equation}}
We note that
QED is at .3\% at both LHC and FNAL, 
this is stable under scale variations,
we agree with the results in Refs.~\cite{baur,ditt,van1,van2},
%BAUR ET AL., HAMBERG ET AL., van NEERVEN and ZIJLSTRA.}\\
and the QED effect similar in size to the structure function 
results~\cite{cern2000,spies,james1,roth,james2}.
DGLAP synthesization~\cite{dglapsyn} has not compromised the normalization.
\par
With the precision tag needed for the LHC in mind, we 
%\titbox{\Color{Maroon}IR-Improved DGLAP Theory}
apply QCD exponentiation theory to DGLAP kernels~\cite{irdglap}: we get{\small
\begin{equation}\begin{split}
P_{qq}(z)= C_F F_{YFS}(\gamma_q)e^{\frac{1}{2}\delta_q}{\big[}\frac{1+z^2}{1-z}(1-z)^{\gamma_q}\cr
 -f_q(\gamma_q)\delta(1-z){\big]}\cr
\label{dglap11}\end{split}
\end{equation}} 
where~{\small
$f_q(\gamma_q)=\frac{2}{\gamma_q}-\frac{2}{\gamma_q+1}+\frac{1}{\gamma_q+2}$,
%\begin{align}
$\gamma_q = C_F\frac{\alpha_s}{\pi}t=\frac{4C_F}{\beta_0}$,
$\delta_q=\frac{\gamma_q}{2}+\frac{\alpha_sC_F}{\pi}(\frac{\pi^2}{3}-\frac{1}{2})$}
%\end{split}
%\label{dglap9}
%\end{align}
%\end{equation}
and~ {\small 
$F_{YFS}(\gamma_q)=\frac{e^{-C_E\gamma_q}}{\Gamma(1+\gamma_q)}$}.
Similar results hold for $P_{Gq},~P_{GG},~P_{qG}$, so that we have:{\small
\begin{align}
%P_{qq}(z)&= C_F F_{YFS}(\gamma_q)e^{\frac{1}{2}\delta_q}\left[\frac{1+z^2}{1-z}(1-z)^{\gamma_q} -f_q(\gamma_q)\delta(1-z)\right],\\
P_{Gq}(z)&= C_F F_{YFS}(\gamma_q)e^{\frac{1}{2}\delta_q}\frac{1+(1-z)^2}{z} z^{\gamma_q},\\
P_{GG}(z)&= 2C_G F_{YFS}(\gamma_G)e^{\frac{1}{2}\delta_G}\{ \frac{1-z}{z}z^{\gamma_G}\nonumber\\
&+\frac{z}{1-z}(1-z)^{\gamma_G}
+\frac{1}{2}(z^{1+\gamma_G}(1-z)\nonumber\\
&+z(1-z)^{1+\gamma_G})- f_G(\gamma_G) \delta(1-z)\},\\
P_{qG}(z)&= F_{YFS}(\gamma_G)e^{\frac{1}{2}\delta_G}\frac{1}{2}\{ z^2(1-z)^{\gamma_G}\nonumber\\
&+(1-z)^2z^{\gamma_G}\},
\label{dglap22}
\end{align}}
%%%Start Here
where~{\small
%\begin{align}
$\gamma_G = C_G\frac{\alpha_s}{\pi}t=\frac{4C_G}{\beta_0}$,
$\delta_G =\frac{\gamma_G}{2}+\frac{\alpha_sC_G}{\pi}(\frac{\pi^2}{3}-\frac{1}{2})$}, and {\small
\begin{align}
f_G(\gamma_G)&=\frac{n_f}{C_G}\frac{1}{(1+\gamma_G)(2+\gamma_G)(3+\gamma_G)}\nonumber\\
&+\frac{2}{\gamma_G(1+\gamma_G)(2+\gamma_G)}\nonumber\\
&+\frac{1}{(1+\gamma_G)(2+\gamma_G)}
%+\frac{1}{12}\}.
+\frac{1}{2(3+\gamma_G)(4+\gamma_G)}\nonumber\\
&+\frac{1}{(2+\gamma_G)(3+\gamma_G)(4+\gamma_G)}.
%\end{split}
\label{dglap19}
\end{align}}

Applying this new kernel set to the evolution of the parton distributions
(Recall that moments of kernels $\Leftrightarrow$ logarithmic exponents for evolution.), 
%we have for the moments of the NS parton distribution
we have for the NS anomalous dimension $A^{NS}_n$ the result
%{\small
%\begin{equation}
%\frac{dM^{NS}_n(t)}{dt}=\frac{\alpha_s(t)}{2\pi}A^{NS}_nM^{NS}_n(t)
%\label{dglap23}
%\end{equation}}
%where{\small
%\begin{equation}
%M^{NS}_n(t)=\int^1_0dz z^{n-1}q^{NS}(z,t)
%\label{dglap24}
%\end{equation}}
%and the quantity $A^{NS}_n$ is given by
{\small
\begin{align}
%A^{NS}_n&=\int^1_0dz z^{n-1}P_{qq}(z),\nonumber\\
A^{NS}_n&= C_F F_{YFS}(\gamma_q)e^{\frac{1}{2}\delta_q}\nonumber\\
&[B(n,\gamma_q)+B(n+2,\gamma_q)-f_q(\gamma_q)]
\label{dglap25}
\end{align}}
where $B(x,y)$ is the beta function
given by {\small$B(x,y)=\Gamma(x)\Gamma(y)/\Gamma(x+y)$.}
Compare the usual result{\small
\begin{equation}
A^{NS^o}_n\equiv C_F [-\frac{1}{2}+\frac{1}{n(n+1)}-2\sum_{j=2}^{n}\frac{1}{j}].
\label{dglap26}
\end{equation}}
%We note that
The IR-improved n-th moment goes for large n to a multiple of $-f_q$,
consistent with $\lim_{n\rightarrow \infty}z^{n-1} = 0$ for $0\le z<1$;
the usual result diverges as $-2C_F\ln n$.
The two results differ for finite n as well: 
we get, for example, for $\alpha_s\cong .118$,
$A^{NS}_2 =C_F(-1.33),C_F(-0.966)$ for (\ref{dglap26}) and (\ref{dglap25}), respectively.
%{\small
%\begin{equation}
%A^{NS}_2 =
%\begin{cases}
%C_F(-1.33)&,~~\text{ un-IR-improved}\\
%C_F(-0.966)&,~~\text{IR-improved}
%\end{cases} 
%\label{dglap27}
%\end{equation}}
For the n-th moment of the NS
parton distribution itself, $M^{NS}_n$, we have~\cite{irdglap} {\small
\begin{equation}
\begin{split}
%%M^{NS}_n(t)&=M^{NS}_n(t_0)e^{\int_{t_0}^{t}dt'\frac{\alpha_s(t')}{2\pi}A^{NS}_n(t')}\cr
M^{NS}_n(t)&=M^{NS}_n(t_0)e^{\bar{a}_n[Ei(\frac{1}{2}\delta_1\alpha_s(t_0))-Ei(\frac{1}{2}\delta_1\alpha_s(t))]} \cr
&\operatornamewithlimits{\Longrightarrow}_{\small t,t_0~\text{large with}~t>>t_0}M^{NS}_n(t_0)\left(\frac{\alpha_s(t_0)}{\alpha_s(t)}\right)^{\bar{a'}_n}
\end{split}
\label{dglap27a} 
\end{equation}}
where $Ei(x)=\int_{-\infty}^xdre^r/r$ is the exponential integral function,
{\small
\begin{equation}
\begin{split}
\bar{a}_n&=\frac{2C_F}{\beta_0}F_{YFS}(\gamma_q)e^{\frac{\gamma_q}{4}}[B(n,\gamma_q)\\
&\qquad +B(n+2,\gamma_q)-f_q(\gamma_q)]\\
\bar{a'}_n&=\bar{a}_n(1+\frac{\delta_1}{2}\frac{(\alpha_s(t_0)-\alpha_s(t))}{\ln(\alpha_s(t_0)/\alpha_s(t))})
\end{split}
\label{dglap27b}
\end{equation}}
with {\small
$\delta_1=\frac{C_F}{\pi}\left(\frac{\pi^2}{3}-\frac{1}{2}\right)$}.
This is to be compared with the un-IR-improved result where last line
in eq.(\ref{dglap27a}) holds exactly with $\bar{a'}_n=2A^{NS^o}_n/\beta_0$.
Comparison with the higher order exact results on the kernels
in Refs.~\cite{mvermn,mvovermn} is done in Ref.~\cite{irdglap}.
%Moch et al., Vogt et al., etc., in progress.}
%\end{itemize}
\par

\section{Final State of Hawking Radiation}
\label{hawkrad}
Consider the graviton propagator in the theory of
gravity coupled to a massive scalar(Higgs) field, as studied
by Feynman in Refs.~\cite{ff1,ff2}. We have the
graphs in Fig.~\ref{fig1} and ~\ref{fig2}.
\begin{figure}
\begin{center}
\epsfig{file=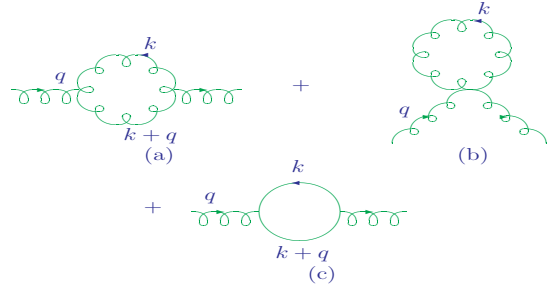,width=77mm,height=38mm}
\end{center}
\caption{\baselineskip=7mm  The graviton((a),(b)) and its ghost((c)) one-loop contributions to the graviton propagator. $q$ is the 4-momentum of the graviton.}
\label{fig1}
\end{figure}
\begin{figure}
\begin{center}
\epsfig{file=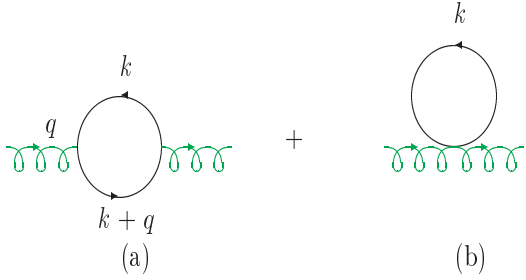,width=77mm,height=38mm}
\end{center}
\caption{\baselineskip=7mm  The scalar one-loop contribution to the
graviton propagator. $q$ is the 4-momentum of the graviton.}
\label{fig2}
\end{figure}
Using the resummed theory, we get that the Newton potential
becomes~\cite{bw1}
\begin{equation}
\Phi_{N}(r)= -\frac{G_NM}{r}(1-e^{-ar}),
\label{newtnrn}
\end{equation}
for $a \cong  0.210 M_{Pl}$.
%%\begin{equation}
%%a \cong  0.210 M_{Pl}.
%%\end{equation}
%
%CONTACT WITH AYMPTOTIC SAFETY APPROACH}
Our results imply{\small
$$G(k)=G_N/(1+\frac{k^2}{a^2})\qquad \qquad $$}
so that we have fixed point behavior for 
$k^2\rightarrow \infty$,
in agreement with the phenomenological asymptotic safety approach of
Refs.~\cite{laut,reuter2,litim}.
%{\Color{Red}BONNANNO \& REUTER IN PRD{\bf 62}(2000) 043008.}\\
We can also see~\cite{bw1} that our results imply 
that an elementary particle has 
no horizon which also agrees with the result of Ref.~\cite{reuter2} 
%{\Color{Red}BONNANNO'S \& REUTER'S\\ 
that a black hole
with a mass less than
 $M_{cr}\sim M_{Pl}$
has no horizon. The basic physics is the following:
 $G(k)$ vanishes as $1/k^2$ for $k^2\rightarrow \infty$.
\par
There is
a further agreement: final state of Hawking radiation of an originally
very massive black hole
because our value of the coefficient, 
$\frac{1}{a^2}$,
of $k^2$ in the denominator of $G(k)$
agrees with that found by
Bonnanno and Reuter(B-R)~\cite{reuter2}, 
if we use their prescription for the
relationship between $k$ and $r$
in the regime where the lapse function
vanishes, we get the same Hawking radiation phenomenology as
they do: the black hole evaporates in the B-R analysis until it reaches a mass
$M_{cr}\sim M_{Pl}$
at which the Bekenstein-Hawking temperature vanishes, 
leaving a Planck scale remnant.
\par
What is the remnant's fate? In Ref.~\cite{bw5}
% hep-ph/0503189
we show that our quantum loop effects combined with
the $G(r)$ of Bonnanno and Reuter imply that the horizon of the Planck scale
remnant is obviated -- consistent with recent results of Hawking~\cite{hawk2}. 
To wit, in the metric class
\begin{equation}
ds^2 = f(r)dt^2-f(r)^{-1}dr^2 - r^2d\Omega^2
\end{equation}
the lapse function is, from Ref.~\cite{reuter2},
\begin{equation}
\begin{split}
f(r)&=1-\frac{2G(r)M}{r}\cr
    &= \frac{B(x)}{B(x)+2x^2}|_{x=\frac{r}{G_NM}},
\end{split}
\end{equation}
where
%\begin{equation}
$B(x)=x^3-2x^2+\Omega x+\gamma \Omega$
%\end{equation}
for
%\begin{equation}
$\Omega=\frac{\tilde\omega}{G_NM^2}=\frac{\tilde\omega M_{Pl}^2}{M^2}$.
%\end{equation}
After Hawking-radiating to regime near $M_{cr}\sim M_{Pl}$, quantum
loops allow us to replace $G(r)$ with $G_N(1-e^{-ar})$
in the lapse function for $r<r_>$, the outermost solution of 
\begin{equation}
G(r)=G_N(1-e^{-ar}).
\end{equation}
In this way, we see that the inner horizon moves to negative $r$ 
and the outer horizon moves to $r=0$ at the new critical mass
$\sim 2.38M_{Pl}$.
We note that the results in Ref.~\cite{bojo} show that
%%M. BOJOWALD {\it et al.}, gr-qc/0503041, 
loop quantum gravity~\cite{lpqg1} concurs with this general conclusion.
\par

We have arrived at the following
prediction: there should energetic cosmic rays at $E\sim M_{Pl}$
due the decay of such a remnant.
\par

%\section{Conclusions}
%\label{conclsn}

%  YFS theory~\cite{yfs,jadward,sjward} (EEX and CEEX) extends to 
%non-Abelian gauge theory and allows simultaneous exponentiation of QED and QCD
%with built-in proper shower/ME matching.
%For QED$\otimes$QCD, full mc event generator realization is possible.
%semi-analytical results for QED (AND QCD) threshold effects agree with literature on Z production. As QED is at the .3\% level, it is needed for 1\% LHC theory predictions. a firm basis for the complete ${\cal O}(\alpha_s^2,\alpha\alpha_s,\alpha^2)$ MC results needed for the FNAL/LHC/RHIC/TESLA/ILC physics 
%has been demonstrated and all the latter is in progress.
%\par
%The theory allows a new approach 
%to quantum general relativity, resummed quantum gravity, which is UV finite.
%There are many consequences: 
%Black holes evaporate to final mass $\sim M_{Pl}$
%with no horizon which implies
%$E\sim M_{Pl}$ cosmic rays, $\cdots$.
%\par
\section*{Acknowledgements}
We thank Profs. S. Jadach and S.A. Yost for useful discussion.
\par
}
\end{document}